# Blockchain e Sistemas Distribuídos: conceitos básicos e implicações

**Maurício Witter**
UFSM
Constantina, BRA
mauricio.witter@ufsm.mail.com

**A. Rodrigo De Vit**
UFSM
Frederico Westphalen, BRA
rodrigodevit@inf.ufsm.br

## Abstract

A tecnologia Blockchain emergiu como uma necessidade da decentralização dos meios de pagamento e transações, mas que trouxe consigo muitas propriedades de sistemas distribuídos que a tornaram uma tecnologia primordial para superar alguns dos desafios da sociedade, especialmente no contexto de decentralização dos serviços, transparência das informações, disponibilidade e segurança. Sua arquitetura e formas de comunicação, embora, possuem algumas nuances complexas de entender, especificamente para o público leigo no assunto de sistemas distribuídos, protocolos e redes de computadores. Neste artigo iremos explorar alguns tópicos de sistemas distribuídos relacionados à tecnologia Blockchain.

## 1 Introduction

A tecnologia *Blockchain* nasceu como uma tecnologia disruptiva que logo conquistou holofotes, principalmente por sua envoltura aplicação em ativos digitais, logo conhecida como o *Bitcoin*. O termo "Blockchain" é usado para descrever uma estrutura de dados, as vezes o sistema como um todo [1, p. 244], mas para este contexto definimos como uma estrutura de dados, onde "Block" se refere a um bloco de transações e "chain" refere-se a cadeia que conecta os blocos por meio de uma *hash*. Assim, a *Blockchain* é uma cadeia de blocos ordenada e encadeada, onde o bloco subsequente contém um *hash* da representação do bloco anterior, como mostra a figura 3.

A tecnologia Blockchain faz uso de uma rede *Peer-to-Peer*[1, p. 243], essa rede é definida como uma rede *overlay* (rede sobreposta). As redes *Peer-to-peer* (P2P) são sistemas distribuídos por natureza, sem qualquer organização hierárquica ou controle centralizado. Os pares formam topologias de redes virtuais sobrepostas auto-organizadas acima da topologia de rede física [2, p. 1]. Essencialmente, os *nodes* da rede formam uma rede virtual que utiliza protocolos gerais para atuar sobre o Protocolo de Internet (IP) e, assim, fazer a conexão entre os pares na rede.

No contexto de arquitetura de software, a tecnologia Blockchain permite novas formas de arquiteturas de software distribuídas, onde o acordo sobre o estado compartilhado para dados descentralizados e transacionais pode ser estabelecido através de uma grande rede de participantes não confiáveis [1, p. 243]. Ou seja, não há um estabelecimento de confiança entre nenhuma das partes, o *node* da rede pode ser um individuo bem-intencionado, uma entidade ou um individuo mal-intencionado, não há o estabelecimento de uma conexão de confiança prévia pois o algoritmo de consenso garante a validade e segurança nas transações, o que é realmente importante para cenários decentralizados.

Este artigo tem como objetivo analisar a arquitetura da tecnologia *Blockchain* e como ela incorpora conceitos de sistemas distribuídos para tornar as aplicações robustas, resilientes, tolerantes a falhas, decentralizadas e seguras. Este trabalho pode ser utilizado futuramente para desenvolver aplicações práticas ou servir como um guia de entendimento a alguns dos conceitos das tecnologias blockchains.

## 2 Peer-to-Peer

Diferentemente do modelo cliente-servidor, que realiza a comunicação diretamente com o modelo TCP/IP, as redes *Peer-to-Peer* são redes virtuais implementadas acima do modelo TCP/IP. A figura 1 apresenta uma conceituação visual de como a rede sobreposta interage com a rede física. Por exemplo, se o *node* de sobreposição **A** tiver tráfego para o *node* **D**, ele poderá roteá-lo diretamente usando o túnel de **A** a **D** ou retransmiti-lo através de outro *node* de sobreposição **B** ou **C** [3].

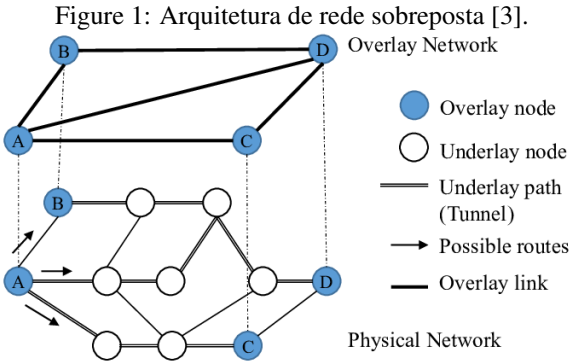

Figure 1: Arquitetura de rede sobreposta [3].

É importante destacar que, a comunicação real entre dois *nodes* conectados por um link na sobreposição é realizada por meio de conexões que existem em uma ou mais camadas de rede subjacentes. A rede física ainda é responsável pelos mecanismos básicos de transporte, endereçamento e roteamento. Os serviços de nível superior como redes de sobreposição utilizam esses mecanismos como meio para a internet [4, p. 17–18].

Enquanto a rede física faz o trabalho de roteamento e transporte para a internet, no nível superior podem ser implementados esquemas de endereçamento e roteamento personalizados para os *nodes* da rede, o que acontece na maioria das aplicações *Peer-to-Peer*, como a *Blockchain*. Assim, sobreposições de alto nível podem ser usadas para desenvolver novos serviços e aplicações sem a necessidade de implantar novos dispositivos ou protocolos na base [4, p. 18].

Figure 2: Topologia Overlay (acima) e Topologia Underlay (abaixo) [4].

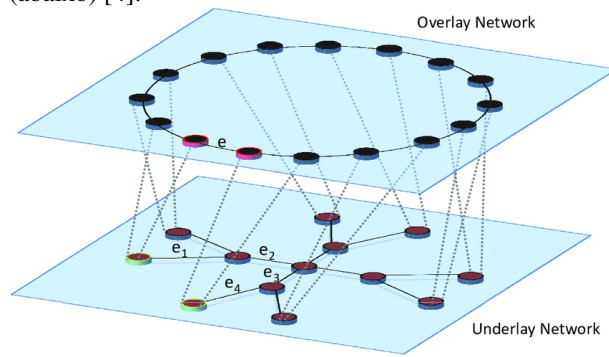

Conforme [2, p. 1], as redes de sobreposição possuem uma combinação de vários recursos, como arquitetura robusta de roteamento de área ampla, pesquisa eficiente de itens de dados, seleção de pares próximos, armazenamento redundante, permanência, nomenclatura hierárquica, confiança e autenticação, anonimato, escalabilidade massiva e tolerância a falhas.

Embora pareça, a seleção de pares não é simplesmente aleatório. Segundo [2, p. 72–73], a topologia da rede de sobreposição P2P é rigidamente controlada e o conteúdo é colocado em locais específicos que farão com que consultas subsequentes sejam mais eficientes. Tais sistemas P2P estruturados usam a Tabela Hash Distribuída (DHT), na qual as informações de localização do objeto de dados (ou valor) são colocadas deterministicamente, nos pares com identificadores correspondentes à chave exclusiva do objeto de dados.

Os sistemas baseados em DHT têm uma propriedade que atribui consistentemente *NodeIDs* aleatórios uniformes ao conjunto de pares em um grande espaço de identificadores. Então, cada par mantém uma pequena tabela de roteamento que consiste nos *NodeIDs* e endereços IP de seus pares vizinhos. Consultas de pesquisa ou roteamento de mensagens são encaminhadas através de caminhos de sobreposição para pares de maneira progressiva, com os *NodeIDs* que estão mais próximos da chave no espaço do identificador [2, p. 73].

Conforme [5, p. 76-79], o Ethereum e o Sistema de Arquivos Interplanetário (IPFS) fazem uso do algoritmo de DHT *Kademlia*. *Kademlia* é um algoritmo de tabela de hash distribuída (DHT) usado em redes *peer-to-peer* (P2P). Este algoritmo é usado principalmente para manter uma lista de *nodes* em uma rede P2P de maneira eficiente e escalável. Ele é conhecido por sua capacidade de pesquisar *nodes* e dados na rede de forma rápida e eficaz. Assim, ele usa *IDs* de *nodes* e *IDs* de recursos para organizar os *nodes* em uma estrutura de árvore binária.

Cada *node* na rede *Kademlia* é identificado por um *ID* único, que é uma sequência binária de tamanho fixo, geralmente de *160 bits*. Os *nodes* são organizados em uma árvore binária, e a proximidade entre dois *nodes* é calculada usando a distância *XOR* entre seus *IDs*. Quanto mais próximos são os *IDs*, mais próximos os *nodes* estão na árvore [2, p. 79–80].

Assim, o uso de redes *overlay* possibilitou novas formas de criar sistemas distribuídos, as novas tecnologias como a Blockchain permitem criar um ecossistema colaborativo e altamente escalável. Os pares compartilham seus recursos computacionais e, além de cliente, torna-se um servidor. Ou seja, cada par da rede atua por uma via bidirecional fazendo ambos os papéis, o que elimina a necessidade de servidores centralizados e contratação de serviços de *cloud computing*.

Cabe ressaltar que, a implementação de redes *overlay* é de alta complexidade, uma vez que não existe de fato uma RFC (*Request for Comments*) ou padronização internacional de implementação, os modelos de endereçamento e roteamento podem ser customizados na rede *overlay*,

o que lhe concede liberdade mas também lhe concerne complexidade.

## 3 Blockchain

A *Blockchain* como um sistema distribuído possui muitos elementos, como já citado, funciona em uma rede *overlay* e utiliza o poder de processamento, a largura de banda e a capacidade de armazenamento das máquinas dos pares da rede. Ademais, possui cadeias e blocos para armazenar os dados com segurança. Mas não obstante, a *Blockchain* é um construto de algoritmos criptográficos, de consenso e de execução de código (*Smart Contracts*).

A primeira geração de Blockchain veio com o Bitcoin [6]. Essa geração elementar possuía muitas limitações que foram sendo aprimoradas ao longo do tempo. Mas, especialmente, a segunda geração de Blockchain trouxe uma revolução ao permitir a implantação e execução de código por usuários em uma Blockchain (*turing-complete*) [1, p. 244].

Uma Blockchain implementa um *ledger* (livro-razão) distribuído, que pode, em geral, verificar e armazenar qualquer tipo de transação (1, p. 1–2 como citado em 7). Cada *node* da rede possui uma cópia idêntica do livro-razão, com isso, os *nodes* da rede podem verificar a partir de algoritmos de consenso as transações de forma transparente e confiável, o que elimina a necessidade de um servidor central confiável.

O livro-razão é o banco de dados de uma *Blockchain* elementar, estruturado pela cadeia de blocos, conforme mostra a figura 3. Esse livro-razão registra todas as transações feitas pelos pares da rede da *Blockchain* e são públicos.

A figura 3 mostra uma representação de como são os blocos de cadeias de uma *Blockchain*. Elementarmente, tem-se uma cadeia linear de blocos, onde o bloco possui seu *hash* e o *hash* do conteúdo anterior.

Figure 3: Estrutura de dados utilizada para representar os blocos e cadeias da Blockchain [8].

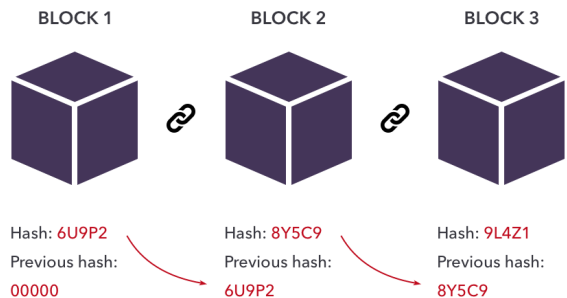

As transações são constituídas de uma carga útil, em outras palavras, pacotes de dados que armazenam parâmetros como valor monetário, endereço do destinatário e resultados de chamadas de função (como contratos inteligentes) [1].

Uma transação passa por uma série de etapas até ser registrada de fato no livro-razão, como mostra a figura 4. Conforme [1], uma transação é assinada por quem a iniciou, para autorizar a carga útil de dados de uma transação ou a criação e execução de um contrato inteligente (*smart contract*).

O remetente assina a transação com sua chave privada (*private key*) para provar que é proprietário da transação. A transação é propagada para os *nodes* conectados à rede *Blockchain*, que fazem a validação inicial a fim de checar se a transação atende as regras do protocolo. Quando a maioria dos pares concorda que a transação é válida, a transação é enviada ao grupo de transações válidas (*mempool*).

Os mineradores da rede (*mining nodes*) fazem a mineração dos blocos através do algoritmo *Proof-of-work* (PoW), quando um *node* cria um bloco, esse bloco é propagado para outros *nodes* na rede, onde também passa por uma validação que requer que a maioria dos pares entre em consenso que é um bloco válido [1, p. 244]. Assim, quando o bloco for considerado válido pelos pares, ele é adicionado à sua cópia da *Blockchain* e propagado aos outros *nodes* da rede.

As transações que estavam no *mempool* são escolhidas pelos mineradores, essa escolha é feita usualmente pela taxa que o remetente pagou, ou seja, quanto maior a taxa, mais rápido a transação será confirmada pelos mineradores. As transações são recolhidas do *mempool* e inseridas nos blocos válidos e propagadas aos pares da rede. Ao final, as transações estão registradas de forma imutável e permanente na *Blockchain*. Conforme, [1], o consenso garante que todas as transações armazenadas sejam válidas e que cada transação válida seja adicionada apenas uma vez.

Figure 4: Etapas que uma transação passa até ser considerada válida e registrada no livro-razão [9].

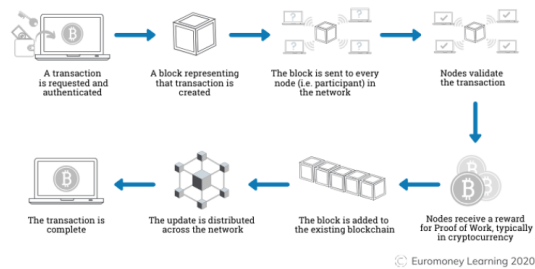

## 4 Segurança

A segurança é outro ponto crucial em Sistemas Distribuídos. A *Blockchain* implementa muitos algoritmos criptográficos para garantir que as transações sejam seguras. Grande parte da segurança é feita pelos algoritmos de consenso, onde todas as transações e blocos precisam passar pela validação da maioria dos *nodes* da rede *blockchain*. De acordo com [1], a integridade de uma transação é verificada por regras algorítmicas e técnicas criptográficas.

O histórico de transações na *Blockchain* não podem ser excluídas ou alteradas sem invalidar a cadeia completa de *hashes*. Combinado com restrições computacionais e esquemas de incentivos à criação de blocos, isso evita a adulteração e revisão das informações armazenadas na *Blockchain* [1, p. 244].

A criptografia de chave pública e as assinaturas digitais são normalmente utilizadas para identificar contas e garantir a autorização de transações iniciadas em uma blockchain [1, p. 244].

Também existem outras técnicas utilizadas em Sistemas Distribuídos que são utilizadas por blockchains, como o *Byzantine fault tolerance* (BFT), uma alternativa ao algoritmo de consenso PoW de [6]. De acordo com [1, p. 250], o BFT exige que todos os participantes concordem com a lista de participantes da rede, por isso, normalmente é utilizado para blockchains privadas, ele é uma abordagem mais convencional em sistemas distribuídos e oferece uma garantia de consistência muito mais forte e menor latência, mas para um número menor de participantes. Conforme [1, p. 250], o BFT garante consenso apesar do comportamento arbitrário de alguma fração dos participantes.

## 5 Consenso

A escolha do protocolo de consenso impacta a segurança e a escalabilidade. Assim que um novo bloco é gerado por um minerador, o minerador propaga o bloco para seus pares conectados na rede *blockchain*. No entanto, os mineradores podem encontrar diferentes novos blocos concorrentes e resolver isso usando os mecanismos de consenso da *blockchain*[1]. A abordagem fundamental proposto por [6, p. 1] foi o algoritmo de consenso *Proof-of-Work* (*PoW*).

No Bitcoin, que utiliza o algoritmo PoW, novos blocos são gerados através do mecanismo de prova de trabalho (*Proof-of-Work*). Os mineradores de Bitcoin competem entre si para resolver cálculos matemáticos simples, mas demorados para decompor, isso é feito para cada bloco, usando grandes quantidades de energia computacional [1, p. 248]. Os blocos, assim como as transações, precisam passar pelo consenso e aprovação da maioria dos pares. Na mineração de blocos, os mineradores concorrem para gerar blocos, os blocos concorrem para se tornarem parte de uma das cadeias de blocos. No Bitcoin, o princípio de consenso é a de que a cadeia mais longa é escolhida, as outras cadeias são abandonadas pelos *nodes* e apenas uma delas é validada. Ao final, o minerador que gerou a bloco vencedor recebe a sua recompensa de mineração somado com as taxas pagas por outros usuários para validarem as transações.

Os sistemas descentralizados que utilizam validadores anônimos precisam de proteção contra ataques *Sybil*, onde os atacantes criam muitos *nodes* anônimos hostis. O *Bitcoin* protege parcialmente contra isso usando seu mecanismo de prova de trabalho, de modo que não é o número total de *nodes* que é importante para a integridade, mas sim a quantidade total de poder computacional. Embora seja fácil para um invasor criar *nodes* anônimos, não é fácil para ele acumular grandes quantidades de poder computacional (1, p. 245 como citado em 10).

## 6 Smart Contracts

*Smart Contracts* (contratos inteligentes) é uma forma de programar contratos (algoritmos) por qualquer pessoa com intuito de executar em uma blockchain *turing-complete* [1, p. 248]. Conforme [11, p. 116675], um contrato inteligente é definido como um programa de computador que faz cumprir as promessas acordadas pelas partes interagentes na ausência de intermediários confiáveis. Assim, a computação em um sistema baseado em blockchain pode ser realizada na cadeia, por exemplo, por meio de contratos inteligentes (*on-chain*) ou fora da cadeia (*off-chain*).

Com o desenvolvimento do ecossistema *Ethereum*, o contrato inteligente se torna um ponto central para alavancar blockchains a máquinas de estado programáveis, introduzindo a execução de aplicativos descentralizados (*dApps*) [11, p. 116675].

Desde a introdução de contratos inteligentes, as aplicações de blockchains não estão mais limitadas à criação e gerenciamento de *tokens* e ativos digitais; surgiram diversas plataformas com recursos de *smart contracts* para conectar blockchains [11, p. 116673].

Contratos inteligentes podem ser desenvolvidos com a linguagem de programação *Solidity*. Esta é uma linguagem *turing-complete* e Orientada a Objetos desenvolvida pela plataforma *Ethereum* para executar contratos inteligentes na Máquina Virtual Ethereum (EVM) [11, p. 116673].

Não é necessário ser um *node* da rede para interagir com a blockchain. Em vez disso, quando um usuário adiciona uma nova entrada ao livro-razão de uma blockchain, ele envia uma transação para um *node* existente usando um protocolo de Chamada de Procedimento Remoto (RPC). Este par retransmite a transação para o resto da rede para inclusão em um bloco futuro. Isto significa que as partes envolvidas numa transação, como o remetente e o destinatário, não estão diretamente envolvidas na execução dessa transação. Em vez disso, esta tarefa cabe aos pares da rede, confirmarem e validarem a transação [12, p. 4]. Assim, os *nodes* da rede fornecem uma interface *JSON-RPC* que permite a qualquer cliente interagir com a blockchain por meio dessa interface de comunicação.

O armazenamento de dados fora da cadeia (*off-chain*) pode ser uma nuvem privada na infraestrutura do cliente ou um armazenamento público fornecido por terceiros ou rede. Alguns armazenamentos de dados *peer-to-peer* são projetados para serem compatíveis com blockchain, como o *IPFS* e o *Storj* [1, p. 248].

Alguns desafios dos contratos inteligentes é que, uma vez implantados à rede principal de uma blockchain, ele não pode mais ser alterado, devido a princípio fundamental de imutabilidade de blockchains.

# 7 Transparência

A transparência é um fator muito importante para sistemas distribuídos. A blockchain como um desses sistemas possui algumas dessas propriedades, tal como a transparência de acesso, onde os participantes têm acesso aos dados e transações registradas na cadeia de blocos por meio de operações padronizadas.

A transparência de concorrência nos algoritmos de consenso como PoW de [6] e PoS permitem transações simultâneas sem interferência entre elas.

A rede *peer-to-peer* é uma forma de transparência de replicação, onde cada *node* da rede possui uma cópia do livro-razão, as transações e inserção de blocos e cadeias são retransmitidas aos *nodes* da rede a todo tempo para garantir que todos tenham uma mesmo cópia, atributos que garantem a disponibilidade e resistência a falhas, assim os usuários finais não precisam se preocupar com a replicação dos dados.

A transparência de falhas é um dos principais problemas de servidores centralizados que a blockchain e redes *peer-to-peer* resolvem, pois mesmo que alguns *nodes* da rede falhem dada uma catástrofe hipotética, há muitos outros *nodes* ativos, o que não irá impactar na indisponibilidade da rede.

Também atende a transparência de mobilidade, os *nodes* da rede podem se conectar e desconectar de forma transparente, sem qualquer impacto aparente a rede *peer-to-peer*.

# 8 Conclusão

As redes *peer-to-peer* tornaram-se muito populares nos últimos anos, devido principalmente a necessidade de descentralizar a web como um todo. O serviços centralizados, apesar de serem maioria, tem algumas desvantagens como a de servidores centralizados, o que implica em complexidades de disponibilidade, em caso de um servidor cair no ponto central, quando não houver meios para recuperação de falhas, o serviço fica *off-line*.

Ademais, conforme [1], em um sistema centralizado, todos os utilizadores dependem de uma autoridade central para mediar as transações. As rede *peer-to-peer*, no entanto, não sofrem com esse problema, uma vez que utilizam os recursos dos *nodes* da rede e podem ficar disponíveis a todo momento e as transações são autônomas e baseadas no consenso. Por exemplo, transações em redes *Blockchain* podem ser feitas a qualquer momento, seja em feriados ou fins de semana, dada sua automação. Já serviços centralizados não-autônomos ficam *off-line* durante períodos eventuais ou rotineiros.

Mas, apesar de todos esses benefícios, há algumas limitações, tal como sistemas em tempo real, tamanho dos dados na *Blockchain*, a latência na confirmação das transações e custos [1]. As *Blockchains* de primeira geração como o *Bitcoin* e o *Ethereum* tem altos custos de transações e as validações podem demorar entre segundos a minutos. Além disso, quanto maior a transação no sentido de tamanho da carga útil (MB/s), maior a taxa cobrada para os *nodes* validarem. Uma prática comum para gerenciamento de dados em sistemas baseados em *blockchain* é armazenar dados brutos fora da cadeia (*off-chain*) e armazenar na cadeia apenas metadados, pequenos dados críticos e *hashes* [1, p. 247].

Alumas *Blockchains* de nova geração como a *Solana* e a *Fantom* possuem outros algoritmos *Proof-of-Stake* (PoS) e estruturas de dados *Directed acyclic graph* (DAG) para organizar e validar as transações de forma mais eficiente sem perder a segurança.

Neste artigos discutimos sobre como conceitos de sistemas distribuídos são aplicados a sistemas reais para fornecer aplicações de alta disponibilidade, distribuídas por meio de redes *overlay* e algoritmos criptográficos para fornecer um sistema seguro para aplicações, onde o desafio está em ter confiança em pares sem conhecimento de boa fé prévia.

Trabalhos futuros podem explorar com maior profundidade os protocolos de comunicação entre redes *overlay*, redes físicas e também explorar as *blockchains* de terceira geração com novas estruturas de livro-razão e algoritmos de consenso, onde chega-se a interoperabilidade entre diversas *blockchains* e governança descentralizada.